\documentclass[prd,aps,12pt]{revtex4}

\usepackage{bm}
\usepackage{amsmath}
\usepackage{yfonts}
\usepackage{epsfig}

\usepackage{graphicx}
\usepackage{rotating}
\usepackage{latexsym}

\def\tt{\tau}

\begin{document}

\newcommand{\beq}{\begin{equation}}
\newcommand{\eeq}{\end{equation}}
\newcommand{\bqa}{\begin{eqnarray}}
\newcommand{\eqa}{\end{eqnarray}}

\begin{flushright}
 BI-TP 2011/05\\
INT-PUB-11-008
\end{flushright}
\vspace{0.8cm}

\thispagestyle{plain}
\setcounter{page}{1}

\title{Does parton saturation at high density explain hadron
multiplicities at LHC ?
}

\author{R.~Baier$^a$\footnote{E-mail address:{baier@physik.uni-bielefeld.de}},
 A.~H.~Mueller$^b$, D.~Schiff$^c$, and D.~T.~Son$^d$}

\affiliation{
$^a$ Fakult\"at f\"ur Physik, Universit\"at Bielefeld, 
D-33501 Bielefeld, Germany\\
$^b$ Department of Physics, Columbia University, New York, NY 10027, USA\\
$^c$ LPT, Universit\'e Paris-Sud, B\^atiment 210, F-91405 Orsay, France\\
$^d$ Institute for Nuclear Theory, University of Washington,
Seattle, WA 98195-1550,  USA}

\vspace{1.1cm}

\begin{abstract}
An addendum to our previous papers in  Phys.~Lett.~{\bf{B539}}~(2002)~46
and Phys.~Lett.~{\bf{B502}}~(2001)~51,  contributed to the CERN meeting "First
data from the LHC heavy ion run", March~4,~2011.
\end{abstract}

\maketitle

\vspace{0.8cm}

This note is an addendum to \cite{Baier:2002bt,Baier:2000sb}. In these papers
we pointed out that charged-particle multiplicities measured in central 
 heavy ion collisions at high energies may not {\sl directly} be determined
by  the initial conditions as given by saturation models, but in addition by
the way gluons are thermalized. This  process may fill the gap between the
multiplicities expected e.g. from saturation models with running coupling BK
equation (rcBK) \cite{Albacete:2010fs} and the ones  measured by the ALICE
Collaboration \cite{Aamodt:2010pb,Aamodt:2010xx} in central Pb-Pb collisions at
 $\sqrt{s_{NN}} = 2.76$ TeV. We argue that this enhancement is evidence
 for a factor $1/\alpha_s^{2/5}$ present in the
weakly-coupled process of the  "bottom-up" equilibration
 mechanism due to nonconservation of the number
of  gluons, i.e. due to entropy production.

\vspace{0.5cm}

At RHIC and LHC energies in the central rapidity region
of heavy ion collisions at high density  energy is deposited
mainly in the form of gluons. 
In saturation  models (hereafter referred to SAT) 
\cite{Kharzeev:2004if,Kharzeev:2007zt,Albacete:2010fs,Levin:2010zy,Armesto:2004ud,Albacete:2010ad,Albacete:2007sm}
 it is  generally assumed 
 that the initial gluon density 
determines the number of produced (charged) particles, i.e. 
that there is a \underbar{direct}
correspondence between the number of partons in the initial state and the 
number of particles in the final state.

In this note we propose a possibly more realistic description, which 
departs from taking the initial condition as the only one determining
 ingredient for the final hadronic state. This description is based on the 
``bottom-up'' scenario \cite{Baier:2002bt,Baier:2000sb}, i.e.
 based on the assumption of 
weakly-coupled processes 
leading to equilibration of a non-Abelian plasma.

Although the original "bottom-up" picture does not account for the physics of
plasma instabilities in the early stages of thermalization \cite{Arnold:2003rq}
which completely change the stages to thermalization except for the last one,
with the result that the thermalization temperature and the thermalization time 
are correctly given by the "bottom-up" estimate \cite{Moore}.
A crucial consequence is that   the number of gluons is
increasing between initial and equilibration times, i.e.
the final  multiplicities
 are not related to the initial condition only,
but also to the way gluons are thermalized.
There is entropy production \cite{Fries:2009wh} here 
dominated by the soft gluons in the final stage
of thermalization \cite{AHM}.
 In the following there is an equality assumed 
between the number of partons in the final state
and the number of observed hadrons ("parton-hadron duality"
\cite{duality}).
\vspace{0.35cm}

Under the SAT assumption the hard gluons (on the saturation scale $Q_s$)
are conserved in number, and they finally
 hadronize, after passing through a hydrodynamical stage.
The "bottom-up" scenario is characterized by the fact that hard gluons 
are degrading, soft ones are formed and start to dominate the system. 
The number of gluons is increasing with proper time $\tt$,
such that the ratio $R$ as the number of soft versus hard initial gluons 
is
 parametrically estimated in terms of the gluon densities
 $n_{soft}(\tau)$ and  
$n_{hard}(\tau)$, respectively,
\begin{equation}
R = [n_{soft}(\tt) (Q_s \tt)] \vert_{      \tt_{eq}}~
/ ~ [n_{hard}(\tt) (Q_s \tt)] \vert_{\tt_0} \sim \alpha_s^{-2/5}  \, ,
\label{ratio}
\end{equation}
derived in  \cite{Baier:2002bt}.

\noindent
In the following, in order to provide explicit  quantitative predictions
for particle multiplicities in this scenario 
we go as much as possible beyond the parametric estimates.
As the main drawback of the model two theoretically undetermined constants
$c_{eq}$ and $c$  have to
be introduced \cite{Baier:2002bt}, i.e. via the equilibration time
\begin{equation}
\tau = \tt_{eq}
= c_{eq}~ \alpha_s^{-13/5} Q_s^{-1} \,
\label{final} \, ,
\end{equation}
and  the equilibration temperature, 
which is expressed by
\begin{equation}
T_{max} = T_{eq} =0.16~ c~c_{eq}~ \alpha_s^{2/5}(Q_s^2)~Q_s  \, .
\label{temp}
\end{equation}

\noindent
In terms of these undetermined parameters
one finds for the ratio $R$,
\begin{equation}
R \simeq  0.13~ c^2~ c_{eq}^4 \alpha_s^{-2/5}(Q_s^2) \, .
\label{rat}
\end{equation}

\noindent
To be quantitative we estimate these two constants with the help of the ALICE
multiplicity measurement \cite{Aamodt:2010pb} together with  RHIC data
at $\sqrt{s_0} = 130 ~GeV$ \cite{Back} as in \cite{Baier:2002bt} in order to
see if plausible estimates for $\tau_{eq}$ and $T_{eq}$ result.

\noindent
The ratio $R (2.76~TeV)$ for the LHC energy at
$\sqrt{s} = 2.76 ~TeV$ is estimated by 
 taking the ALICE value \cite{Aamodt:2010pb} 
\begin{equation}
 dN_{ch}/d\eta = 1584 \pm 4 (stat) \pm 76 (sys) \, ,
\label{Alice}
\end{equation}
and as a reference value for SAT the one from \cite{Albacete:2010fs},

\begin{equation}
 dN_{ch}/d\eta ~(2.76 ~TeV) = 1175 \pm 75 \, ,        
\label{Albacete}
\end{equation}
which underpredicts the ALICE value (see FIG. ~4 in \cite{Aamodt:2010pb}),
as
\cite{Albacete:2010fs}  relates the multiplicity directly to the CGC initial
 state configuration, and it solves the BK equation using the
running coupling (rcBK) \cite{Albacete:2007sm}.

\noindent
The underestimated value (\ref{Albacete}) may be enhanced
  by the interaction of gluons, which are
redistributed and thermalized. 
Using (\ref{Alice}) and (\ref{Albacete}) we estimate
\begin{equation}
 R (2.76 ~TeV) \simeq 1.35 ~(\pm 0.15)      \, .
\label{estimate}
\end{equation}

\noindent
We take  the parametrizations from \cite{Baier:2002bt}
at $ \sqrt{s_0} = 130~ GeV$:
\begin{equation}
 Q_s^2 = (s/s_0)^{\lambda/2} ~ Q_{s0}^2,~~  \lambda = 0.25, ~~
Q_{s0}^2 = 2 ~GeV^2,  ~ \alpha_s = 0.35 \, ,
\label{param1}
\end{equation}
and
\begin{equation}
   Q_s^2 ~(2.76) = 4.3~ GeV^2,~~ \alpha_s = 0.3~ ~,
\label{param2}
\end{equation}
with $\Lambda_{QCD} = 0.2~GeV$.

\noindent
 Although there is a lot of freedom,  as an illustrative numerical example
we choose $R(130~GeV) c = 3$ as in \cite{Baier:2002bt}
 to fix the RHIC multiplicity \cite{Back}. Plausible $O(1)$ estimates for the constants
follow:
\begin{equation}
        c \simeq 2.4 ~(\pm 0.2), ~~~
        c_{eq} \simeq 1.1 ~(\pm 0.1) \, .
\label{param3}
\end{equation}

\noindent
Eqs.~(\ref{final}) and (\ref{temp}) 
 give the estimates  $\tau_{eq} \simeq 2.2~ fm$
and   $T_{eq} \simeq 490~ MeV$ 
at the LHC energy $\sqrt{s} = 2.76~TeV$,
to be reasonable values   in the framework of pQCD.

\noindent
For the higher energy at $\sqrt{s}= 5.5~ TeV$ we expect, using
$Q_s^2 = 5.1~ GeV^2,~ \alpha_s = 0.29 , ~R(5.5~TeV) \simeq 1.38$,
\begin{equation}
   dN_{ch}/d\eta \simeq 1910~ (\pm 50) \, ,
\end{equation}
enhancing  the SAT value \cite{Albacete:2010fs}
\begin{equation}
dN_{ch}/d\eta = 1390 \pm 95 \, .
\end{equation}

\noindent
Using $N_{part} = 381$ (taken at $ 2.76~ TeV$ \cite{Aamodt:2010pb})
gives
\begin{equation}
\frac{2}{N_{part}} dN_{ch}/d\eta \simeq  10.0 \, ,
\end{equation}
in agreement with the ALICE fit  using the proportionality 
$ \propto{ s^{0.15}}$ (see FIG.~3 \cite{Aamodt:2010pb}).

\noindent
The dependence of $d N_{ch}/ d\eta$ as a function of $N_{part}$ found by the 
ALICE Collaboration (see FIG.~2 in \cite{Aamodt:2010xx}) is similar to that observed at RHIC 
energies \cite{Back,Back:2001xy,Adler:2004zn,Abelev:2008ez}
 which has been shown to be consistent with the "bottom-up" expectation 
(see Fig.~2 in \cite{Baier:2002bt}).

\vspace{0.5cm}
In summary, the description for particle multiplicities 
provided by the ``bottom-up'' scenario, due to inherent entropy production, 
reproduces the RHIC  and LHC data,
provided the parameters $c$ and $c_{eq}$, which are not determined
in the picture, lie in a given, limited
range of $O(1)$.
Due to the interactions of gluons, which are redistributed and thermalized
 in this kinetic scenario,
 the initial gluon spectrum is strongly modified, i.e. enhanced.


\begin{thebibliography}{99}


\bibitem{Baier:2002bt}
  R.~Baier, A.~H.~Mueller, D.~Schiff and D.~T.~Son,
  ``Does parton saturation at high density explain hadron multiplicities 
at RHIC?,''
  Phys.\ Lett.\  {\bf B539 } (2002)  46-52.
  [hep-ph/0204211].

\bibitem{Baier:2000sb}
  R.~Baier, A.~H.~Mueller, D.~Schiff and D.~T.~Son,
  ``'Bottom up' thermalization in heavy ion collisions,''
  Phys.\ Lett.\  {\bf B502 } (2001)  51-58.
  [hep-ph/0009237].


\bibitem{Albacete:2010fs}
  J.~L.~Albacete,
  ``CGC and initial state effects in Heavy Ion Collisions,''
  [arXiv:1010.6027 [hep-ph]].

\bibitem{Aamodt:2010pb}
  K.~Aamodt {\it et al.} [ The ALICE Collaboration ],
  ``Charged-particle multiplicity density at mid-rapidity in central
 Pb-Pb collisions at sqrt(sNN) = 2.76 TeV,''
 Phys.\ Rev.\ Lett.\  {\bf 105 } (2010)  252301.  
  [arXiv:1011.3916 [nucl-ex]].


\bibitem{Aamodt:2010xx}
  K.~Aamodt {\it et al.} [ The ALICE Collaboration ],
  ``Centrality dependence of the 
charged-particle multiplicity density at mid-rapidity in 
 Pb-Pb collisions at sqrt(sNN) = 2.76 TeV,''
  Phys.\ Rev.\ Lett.\  {\bf 106 } (2011)  032301. 
  [arXiv:1012.1657 [nucl-ex]].


\bibitem{Kharzeev:2007zt}
  D.~Kharzeev, E.~Levin, M.~Nardi,
  ``Hadron multiplicities at the LHC,''
%
  [arXiv:0707.0811 [hep-ph]].

\bibitem{Kharzeev:2004if}
  D.~Kharzeev, E.~Levin, M.~Nardi,
  ``Color glass condensate at the LHC: Hadron multiplicities 
in pp, pA and AA collisions,''
  Nucl.\ Phys.\  {\bf A747 } (2005)  609-629.
  [hep-ph/0408050].



\bibitem{Levin:2010zy}
  E.~Levin, A.~H.~Rezaeian,
  ``Hadron multiplicity in pp and AA collisions at LHC from
 the Color Glass Condensate,''
  Phys.\ Rev.\  {\bf D82 } (2010)  054003.
  [arXiv:1007.2430 [hep-ph]].


\bibitem{Armesto:2004ud}
  N.~Armesto, C.~A.~Salgado, U.~A.~Wiedemann,
  ''Relating high-energy lepton-hadron, proton-nucleus
 and nucleus-nucleus collisions through geometric scaling,''
  Phys.\ Rev.\ Lett.\  {\bf 94 } (2005)  022002.
  [hep-ph/0407018].


\bibitem{Albacete:2010ad}
  J.~L.~Albacete, A.~Dumitru,
  ``A model for gluon production in heavy-ion collisions 
at the LHC with rcBK unintegrated gluon densities,''
  [arXiv:1011.5161 [hep-ph]].

\bibitem{Albacete:2007sm}
  J.~L.~Albacete,
  ``Particle multiplicities in Lead-Lead collisions
 at the LHC from non-linear evolution with running coupling,''
  Phys.\ Rev.\ Lett.\  {\bf 99 } (2007)  262301.
  [arXiv:0707.2545 [hep-ph]], and references therein.

\bibitem{Arnold:2003rq}
  P.~B.~Arnold, J.~Lenaghan, G.~D.~Moore,
  ``QCD plasma instabilities and bottom up thermalization,''
  JHEP {\bf 0308 } (2003)  002.
  [hep-ph/0307325].


\bibitem{Moore}
G.~Moore, private communication.

\bibitem{Fries:2009wh}
For an independent discussion, see:
  R.~J.~Fries, T.~Kunihiro, B.~Muller, A.~Ohnishi, A.~Schafer,
  ``From 0 to 5000 in 2 x 10**-24 seconds:
 Entropy production in relativistic heavy-ion collisions,''
  Nucl.\ Phys.\  {\bf A830 } (2009)  519C-522C.
  [arXiv:0906.5293 [nucl-th]].




\bibitem{AHM}
A.~H.~Mueller,
``Equilibration in very high energy heavy ion collsions,''
Nucl. Phys. {\bf{A702}} (2002) 65c.

\bibitem{duality}
Yu.~L.~Dokshitzer, V.~A.~Khoze, A.~H.~Mueller and S.~I.~Troyan,
{\sl Basics of perturbative QCD}, \\
 Editions Fronti\'ers, Gif-sur-Yvette, 1991, and references therein.


\bibitem{Back}
  B.~B.~Back {\it et al.} [ PHOBOS Collaboration ],
  ``Charged particle multiplicity near mid-rapidity in central Au + Au collisions at S**(1/2) = 56-A/GeV and 130-A/GeV,''
  Phys.\ Rev.\ Lett.\  {\bf 85 } (2000)  3100-3104.
  [hep-ex/0007036].

\bibitem{Back:2001xy}
  B.~B.~Back {\it et al.} [ PHOBOS Collaboration ],
 ``Centrality dependence of charged particle multiplicity at mid-rapidity in Au + Au collisions at s(NN)**(1/2) = 130-GeV,''
  Phys.\ Rev.\  {\bf C65 } (2002)  031901.
  [nucl-ex/0105011].

\bibitem{Adler:2004zn}
  S.~S.~Adler {\it et al.} [ PHENIX Collaboration ],
  ``Systematic studies of the centrality and s(NN)**(1/2) dependence
 of the $ d E(T) / d \eta$ and $d (N(ch) / d \eta$ in heavy ion collisions at mid-rapidity,''
  Phys.\ Rev.\  {\bf C71 } (2005)  034908.
  [nucl-ex/0409015].

\bibitem{Abelev:2008ez}
  B.~I.~Abelev {\it et al.} [ STAR Collaboration ],
  ``Systematic Measurements of Identified Particle Spectra in pp, d+Au and Au+Au Collisions from STAR,''
  Phys.\ Rev.\  {\bf C79 } (2009)  034909.
  [arXiv:0808.2041 [nucl-ex]].



\end{thebibliography}
\end{document}